\documentclass[preprint2]{aastex}
\usepackage{spr-astr-addons}                                  
\usepackage{url}

\RequirePackage{color}

\begin{document}

\title{Optical flare activity in the low-mass eclipsing binary GJ~3236}

\shorttitle{Flare activity in GJ~3236}
\shortauthors{Parimucha, Dubovsk\'y, Va\v{n}ko \& \v{C}okina}

\author{\v{S}. Parimucha\altaffilmark{1}} 
\affil{Institute of Physics, Faculty of Science, University of P.J. \v{S}af\'{a}rik, 044~01 Ko\v{s}ice, 
Slovakia }
\email{stefan.parimucha@upjs.sk}

\and 

\author{P. Dubovsk\'y\altaffilmark{2}}
\affil{Vihorlat Observatory in Humenn\'e, 066~01 Humenn\'e, Slovakia}

\and 

\author{M. Va\v{n}ko\altaffilmark{3}} 
\affil{Astronomical Institute, Slovak Academy of Sciences, 059~60 Tatransk\'a
Lomnica, Slovakia}

\and 

\author{M. \v{C}okina\altaffilmark{1}} 
\affil{Institute of Physics, Faculty of Science, University of P.J. \v{S}af\'{a}rik, 044~01 Ko\v{s}ice, 
Slovakia }

\maketitle

\begin{abstract}
We present our observations of the low-mass eclipsing binary GJ~3236. We have analyzed a phased $R_C$ light-curve and 
confirmed previously determined fundamental parameters of the components. We detected evolution of the spot(s) and 
found that there exists a large spot near a polar region of the primary component and another spot either on the 
primary or the secondary component. We also observed 7 flare events and determined a flare rate of about 0.1 flares per 
hour. We observed two high energy, long-term flares with a complex light curve and possibly four weak short-term 
flaring events. A majority of the flares was detected in the $R_C$ filter, which indicate their high energy.
\end{abstract}

\keywords{binaries: eclipsing -- binaries: low-mass -- stars: flares -- stars: activity -- stars: spots}

\section{Introduction}
\label{s:1}

In what follows, the term $flare$ will refer to violent events in stellar atmospheres, which release accumulated 
magnetic energy and reject hot material into space. We often observe them on magnetically active stars, and of course, 
on the Sun. The radiation is emitted across the whole spectrum, from $\gamma$-rays to radio waves. The total 
bolometric energy released during such a flare event can amount to 10$^{24}$ - 10$^{27}$~J \citep{pettersen1989}. 

Flares can be modeled by a beam of charged particles that are accelerated by oppositely-oriented magnetic field lines 
at upper levels of a stellar atmosphere. They cause acceleration of particles in the beam and propagate downward 
into the star's surface. During this, particles deposit energy and momentum along the way and produce electromagnetic 
radiation. This rapidly heats the lower stellar atmosphere, causing it to explosively expand and dramatically brighten 
\citep{allred2015}. Stellar flares, however, cannot be explained by a simple scaling of solar-like flares to stars. 
Observations uncover problems  arising from the different spectral distribution of the emitted energy, the role of 
the age and spectral type of the host star, the multiform magnetic field topologies on stars, the tidal forces in 
active binaries, etc. \citep{kovari2007}. 

Flare activity is a very common phenomenon in so-called red dwarfs - low mass, low luminous and cool stars
\citep{davenport2012}. This  group of stars is the most abundant group in our galaxy. Up to date there are known only 
about twenty eclipsing binaries where both components are red dwarfs, because this low luminosity systems are quite 
difficult to uncover. Using high dispersion spectroscopy it is possible to determine basic parameters of the components 
with an accuracy of about 1\%. This is sufficient for evolutional modeling of these stars, which also play a key role 
in our understanding of stellar physics of main sequence stars at the bottom part of the HR diagram. Massive convection 
affects the significant part of the atmosphere of red dwarfs and generates strong magnetic fields, the source of high 
activity which manifests itself through intensive flares probably originating in the chromosphere and the corona of red 
dwarfs. Strong magnetic activity is also manifested by the presence of spots on the surface of the stars, which 
strongly affects the shape of light curves of eclipsing binaries containing theses stars \citep{irwin2009, Ribas2003}.

In this paper we present our light curve analysis and detection of optical flares in such a eclipsing system GJ~3236 
(LSPM J0337+6910, 2MASS J03371407+6910498, USNO-B1.0 1591-0058025, $\alpha_{2000}$ = 03$^h37^m14^s.08$, $\delta_{2000}$ 
= +69\degr 10' 49''.8, $V$=12.21 mag, $I$=10.9 mag). This system was for the first time noted by \cite{gliese1991} in 
the Third Catalogue of Nearby Stars. \cite{hawley1996} included this star into a spectroscopic survey of M class dwarfs 
in the northern hemisphere and \cite{hunsh1999} included GJ~3236 in the catalog of nearby stars observed by the ROSAT 
satellite. Since then it has been listed also as an X-ray source. Eclipses on the light curve were discovered by 
\cite{irwin2009}, who established the  ephemeris of the binary system to be 
\begin{equation}
 HJD_{min}=2454734.99586(12) + 0^{d}.7712600(23).
 \label{eq:ephemeris}
\end{equation}
From radial velocity measurements of the both components the authors derived their physical properties and found that 
the system is binary with red dwarfs components, of the spectral type dM. The system shows signs of high activity 
and evidence of strong $H\alpha$ emission lines in the spectra of both components. \cite{dittmann2014} determined 
trigonometric paralax for GJ~3236 to be 27.7$\pm$1.3 marcsec, what corresponds to distance 36.1$\pm$1.7 pc.
The optical flare activity in this binary was for the first time reported by \cite{smelcer2015}.

\section{Observations}

Presented photometric observations were carried out at the Astronomical Observatory at the Kolonica Saddle 
\citep{kudzej2007}. For all observations except the last one (March, 19. 2016) we used the Celestron C14 Edge HD 
telescope (aperture 356 mm, f/11) with a Moravian Instruments G2-1600 CCD camera (1536$\times$1024 pixels). This 
configuration gives us the field of view of 12' $\times$ 8' with the resolution of 0.84''/pixel (with 2$\times$2 
binning). The last observation was performed by the \v{Z}IGA telescope operated by the Institute of Physics, University 
of P. J. \v{S}af\'{a}rik, at the same observatory. It is a Planewave CDK20 telescope (aperture 508mm f/6.8) with 
a Moravian Instruments G4-16000 CCD camera (4096$\times$4096 pixels). This configuration gives us the field of view 
of 36'$\times$ 36' with the resolution of 1.07''/pixel (with 2$\times$2 binning).

The CCD images were reduced in the usual way (bias and dark subtraction, flat-field correction) and aperture 
photometry was performed using the software package C-Munipack\footnote{\tt http://c-munipack.sourceforge.net/}. 
The first-order nightly atmospheric extinction coefficients were computed from the comparison stars' measurements and 
the correction for the atmospheric extinction was applied to these observations. 

To improve our photometric precision and eliminate possible variability of the comparison star(s), we used a multiple 
comparison star method \citep{kim2004} as implemented in the package MCV \citep{andronov2004} and computed instrumental 
differential magnitudes according to the main comparison star  GSC~4327-0784, ($V$=12.67, $B$=13.49, $R$=12.05, 
$I$=11.47) \citep{zacharias2015}. We did not perform  any correction to the international system, because a majority 
of our observations were carried out only in one photometric passband. But we made night-to-night systematic 
corrections using the TFA method \citep{kovacs2005} implemented in the VARTOOLS Light Curve Analysis Program 
\citep{hartman2008}. The average accuracy of our measurements achieved 0.018 mag in the $V$ filter, 0.007 mag in $R_C$ 
and 0.006 mag in $I_C$ filters, respectively.

A list of our observations of the GJ~3236 system used in this study, with basic information, about each night, is 
presented in Table~\ref{tbl:2}. 

\begin{table}[t]
\caption{The summary of our observations of the eclipsing binary GJ~3236 used in this study. The phase is given 
according to ephemeris (1).} 
\label{tbl:2}
\begin{center}
  \begin{tabular}{lccc}
   \tableline  
   Date     &  Filters  & Exp. time[s] & Phase  \\
   \tableline  
 2015-10-30 &     $R_C$ &  60          &  0.701 - 0.336\\
 2015-10-31 &     $R_C$ &  60          &  0.987 - 0.607\\  
 2015-11-01 &     $R_C$ &  60          &  0.297 - 0.579\\  
 2015-11-03 &     $R_C$ &  60          &  0.872 - 0.254\\  
 2015-11-05 &     $R_C$ &  60          &  0.699 - 0.112\\  
 2015-11-24 &     $V  $ &  60          &  0.128 - 0.451\\  
            &     $I_C$ &  30          &  0.137 - 0.537\\  
 2015-12-29 &     $V  $ &  60          &  0.479 - 0.041\\          
            &     $R_C$ &  30          &  0.480 - 0.040\\          
 2016-02-05 &     $V  $ &  60          &  0.869 - 0.095\\          
            &     $R_C$ &  30          &  0.867 - 0.096\\ 
 2016-03-19 &     $V  $ &  30          &  0.602 - 0.921\\          
            &     $R_C$ &  20          &  0.603 - 0.920\\             
  \tableline  
   \end{tabular}
\end{center}
\end{table}

\begin{table}[t]
\label{tbl:parameters}
\caption{Photometric parameters of GJ~3236 and their statistical $1\sigma$ uncertainties obtained from two light-curve 
solutions (see the text). Values denoted by the superscript $^f$ are fixed during solutions. $T_{1,2}$ - 
the temperatures of the components, $\Omega_{1,2}$ - their surface potentials,  $i$ - the orbital inclination, $q$ - 
the mass ratio, $\theta$, $\phi$, $r$, $k$ - the longitude, latitude, radius and temperature factor of the spots,   
$\sum w \chi^2$ - the weighted sum of squares of residuals.} 
\begin{center}
  \begin{tabular}{lcc}
   \tableline  
   Parameter          &  Solution 1  & Solution 2 \\
   \tableline  
   $T_1$[K]           &    3341$\pm$86     &    3346$\pm$89  \\  
   $T_2$[K]           &    3235$\pm$86     &    3241$\pm$82  \\  
   $\Omega_1$         &    8.68$\pm$0.12   &    8.71$\pm$0.11  \\  
   $\Omega_2$         &    9.73$\pm$0.11   &    9.74$\pm$0.11 \\  
   $i$[\degr]         &     83.1$\pm$0.3   &    83.1$\pm$0.3 \\  
   $q$                &\multicolumn{2}{c}{0.746$^f$}  \\
   \tableline
   Spots:      &        &      \\  
   \tableline
   $\theta_{P,1}$[\degr]     &      107$\pm$14  &    109$\pm$13  \\  
   $\phi_{P,1}$  [\degr]     &      14$\pm$7    &    15$\pm$8  \\  
   $r_{P,1}$     [\degr]     &      62$\pm$19   &    66$\pm$20  \\  
   $k_{P,1}$                 &    0.9$^f$   &    0.9$^f$ \\  
   $\theta_{P,2}$[\degr]     &   --         &    11$\pm$11  \\  
   $\phi_{P,2}$  [\degr]     &   --         &    54$\pm$9  \\  
   $r_{P,2}$     [\degr]     &   --         &    27$\pm$16  \\  
   $k_{P,2}$          &          --         &    0.9$^f$  \\  
   $\theta_{S,1}$[\degr]     &   184$\pm$15     &   --  \\  
   $\phi_{S,1}$  [\degr]     &   92$\pm$7      &   --  \\  
   $r_{S,1}$     [\degr]     &   44$\pm$21     &   --  \\  
   $k_{S,1}$                 &    0.9$^f$   &    --  \\  
   \tableline
   $\sum w \chi^2$             &    0.1538    &    0.1502 \\         
  \tableline  
   \end{tabular}
\end{center}
\end{table}

\section{Light curve of GJ~3236}

The first, and up to now only one, light curve solution of GJ~3236 was given by \cite{irwin2009}, who analyzed 
photometric observations from the MEarth project \citep{nutzman2008}, together with their own radial velocity 
measurements of the both components (see Figures 2 and 3 in their paper). They found that GJ~3236 is a low-mass binary 
with the component masses of 0.38$M_\odot$ and 0.28$M_\odot$, and with the stellar radii of 0.38$R_\odot$ and 
0.30$R_\odot$, which are about 10\% larger than the theoretical  predictions. This is in agreement with the
results obtained for known low-mass binaries and is mostly explained by large areas of dark spots on the surfaces 
of both components, which decrease the stars' radiating surfaces and also yield a smaller effective temperature 
and a larger radius \citep{Chabrier2007}. \cite{irwin2009} modeled their light curve by cool, close to polar spots on 
the surface of each component. 

\begin{figure}[t]
 \includegraphics[width=\columnwidth]{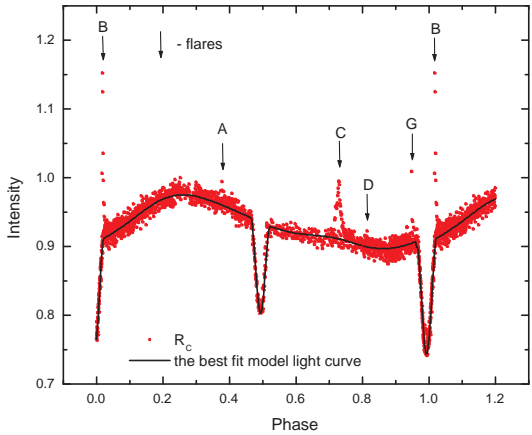}  
\caption{The phase light curve of GJ~3236 obtained from our $R_C$ observations according to ephemeris (1). We 
also denote detected flares in the $R_C$ passband (see Table \ref{tbl:3}) and give the best-fit-model light curve 
for Solution 2 (see the text).}
\label{fig:01}
\end{figure}

Our $R_C$ light curve (see Fig~\ref{fig:01}.) is markedly different from the MEarth ones and cannot be fitted by the 
model from \cite{irwin2009}. A detrended light curve is quite stable during the interval of our observations up to 
changes (about 0.03 mag) in the phase interval 0.5 -- 1.0. In this interval two large flares occurred (see Section 
\ref{flare_activity}), so we concluded that these night-to-night variations are connected with flare activity.

We removed flare events from the phased light curve (see Section \ref{flare}) and analyzed it with the PHOEBE package 
\citep{Prsa2005}. As initial parameters for the components temperatures, surface potentials and orbital inclination 
we adopted the values from \cite{irwin2009}. The mass ratio $q$ was fixed to be 0.746 as a result of radial velocity 
solution. 

The observed variations of the light curve can be explained by the presence of the spots. We have tested several 
different configurations of spots' locations, dimensions and temperatures. The light curve shape in the phase interval 
of 0.0--0.5 at the first sight suggests presence of a hot spot on the surface of one of the components. But the 
solutions with hot spot(s) on the primary and/or secondary component in combination with one or more cool spots 
were not sufficient, mainly because of the inadequate depth of the primary minimum, although the secondary minimum and 
out of minima part were fitted quite well. So we reject this solution as not being a proper one.

To explain the depth of the primary minimum, as well as variations on the light curve, we have two possible solutions. 
If we assume that the primary minimum is caused by the passage of secondary (smaller and less massive) star in front of 
the primary (large, more massive) star, we need to reduce the light from system during this passage. This could be done 
by: 
\begin{enumerate}
	\item a cool spot placed on the secondary component on the hemisphere visible during the minimum and another 
		spot close to the polar region of the primary component visible mainly in the phase interval of
		0.5--1.0.
	\item two cool spots on the primary component, one close to the polar region as in the previous case, 
		and the second one visible during the primary minimum.
\end{enumerate}

\begin{figure}[t]
\begin{center}
 \includegraphics[width=0.7\columnwidth]{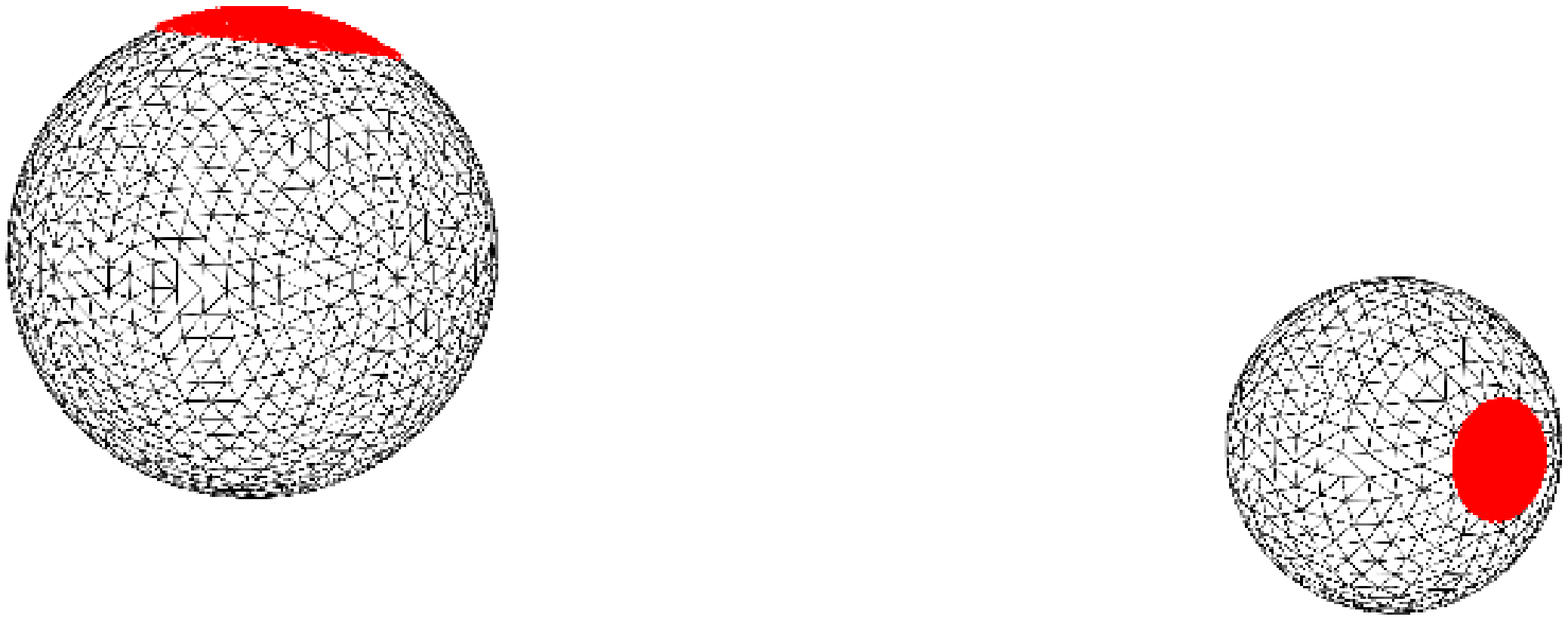}  
 \includegraphics[width=0.7\columnwidth]{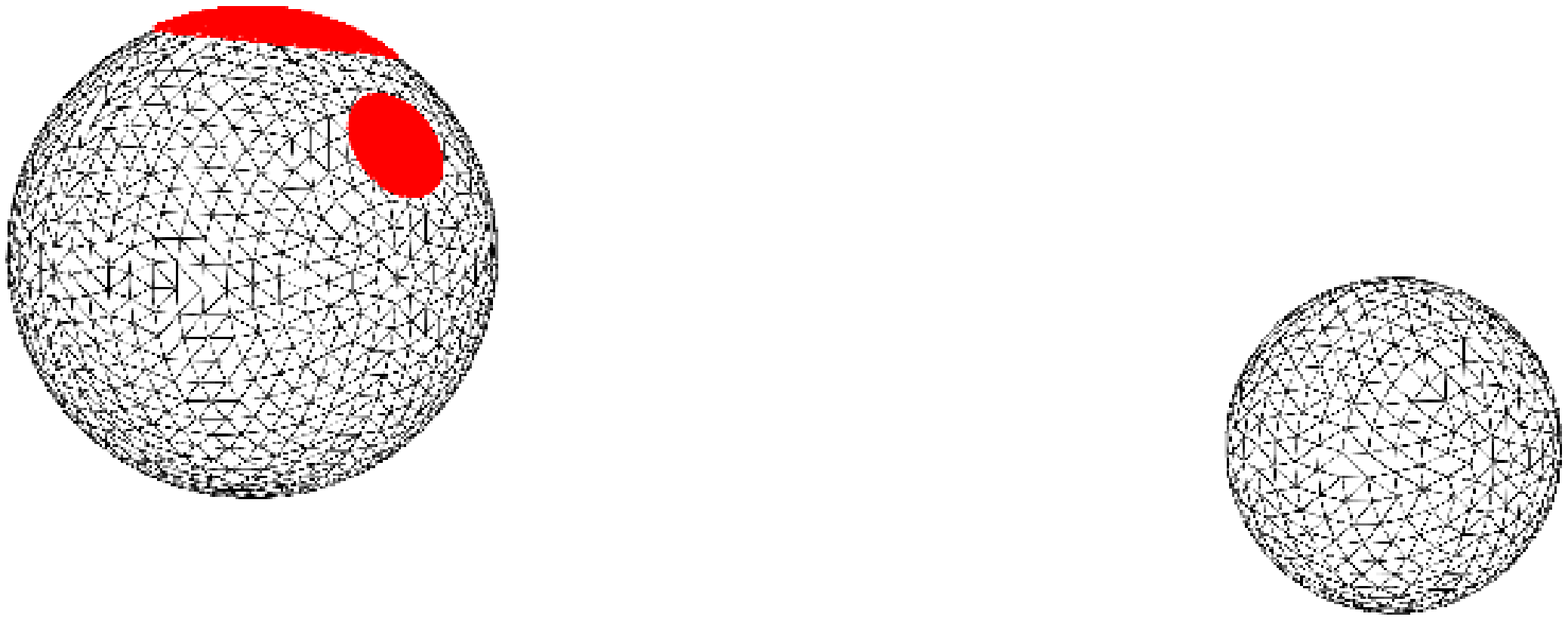} 
\end{center}
 \caption{A model of components in GJ~3236 with spots' locations (see the text) for Solution 1 (top) and Solution 2 
(bottom).}
\label{fig:02}
\end{figure}

During solution of the light curve, we fixed the temperature factor of the spots and varied their positions (longitude 
and latitude) and radii. We also varied components' temperatures, surface potentials and orbital inclination. These 
values did not change significantly from the solution of \cite{irwin2009} and did not have strong influence on the 
model light curve. 

The PHOEBE package gives formal statistical errors of the fitted parameters determined from the covariance matrix.
These values are often underestimated. To obtain more realistic robust statistical uncertainties of the 
fitted parameters we used the Monte Carlo simulation method \citep{press2002}. We assumed that the probability 
distribution of our measurements' errors is normal (Gaussian). We produced 10000 synthetic LCs around the best solution 
and fitted them using capabilities of the PHOEBE-Scripter to obtained set of parameters and corresponding $\chi^2$. 
Statistical errors of individual parameters were established as a $1\sigma$ confidence interval of the projection of an
11-dimensional parameters space.

In Table~\ref{tbl:parameters} we present results for both spotted solutions. Our results show that 
Solution 2 has a smaller weighted sum of squares of residuals, but without other type of observations 
(e.g. high resolution spectroscopic Doppler imaging) we cannot conclude which model is correct. Our photometric 
solution did not markedly change determination of absolute parameters of the components from \cite{irwin2009} and all 
values can be considered equal within the errors.

The phase light curve of GJ~3263, together with the best fit model light curve for Solution 2, is depicted in 
Fig~\ref{fig:01}. Fig~\ref{fig:02} shows the configuration of the components with spots locations listed in 
Table~\ref{tbl:2}.

\label{flare_activity}
\begin{figure}[t]
 \includegraphics[width=\columnwidth]{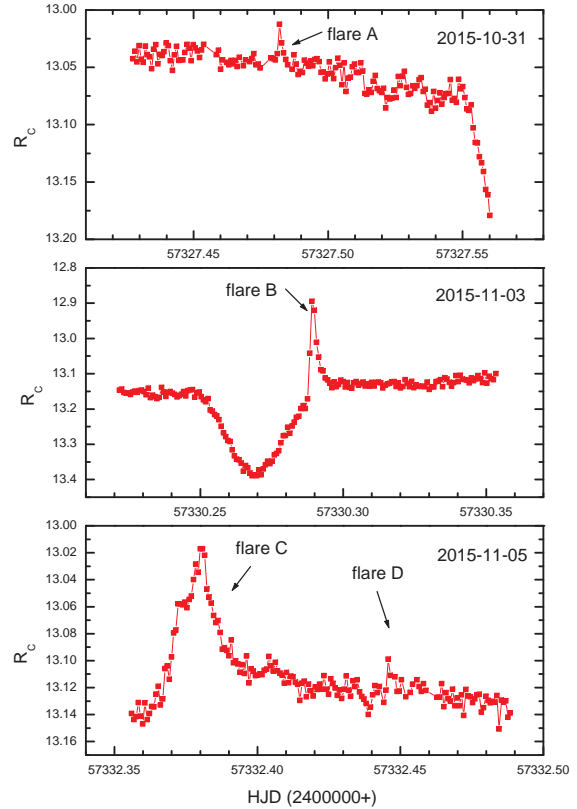}  
\caption{Flares detected in the $R_C$ passband at different nights. For the parameters of the flares, see 
Table~\ref{tbl:3}.}
\label{fig:03}
\end{figure}

\section{Flare activity}
\label{flare}

\begin{table*}[t]
\caption{Flares detected in GJ~3236 and their parameters: Label - designation of the flare, Date - the date of 
observation, Start, End - estimates when the flare started or finished, Maximum - the time of maximum brightness, 
Maximum phase - the phase according to ephemeris (1) of maximum brightness, $\Delta$m - the brightening amplitude, 
Duration - the duration of the flare, Energy - estimated energy with the error in parenthesis (for discussion about 
energy see Section \ref{disscussion}).} 
\label{tbl:3}
\begin{center}
  \small
  \begin{tabular}{clccccccc}
   \tableline  
 Label  &Date     &  Start     & Maximum     & End     &  Maximum &$\Delta$m& Duration    & Energy\\
        &    &  \multicolumn{3}{c}{HJD}            &  phase        &   &  [min]      &    [10$^{25}$J]\\
   \tableline  
 A  &2015-10-31&  57327.4799 &  57327.4819    &  57327.4841 &  0.378 &0.04 ($R_C$)& 6.1  &1.1(7)\\  
 B  &2015-11-03&   --        &  57330.2891    &  57330.2972 &  0.023 &0.24 ($R_C$)& $>$10 &$>$4.8(5)\\  
 C &2015-11-05& 57332.3663  &  57332.3803    &  57330.3987 &  0.729 &0.10 ($R_C$)& 46.6  &5.7(5)\\  
 D &	    & 57332.4447  &  57332.4457    &  57330.4494 &  0.814 &0.03 ($R_C$)& 6.7   &0.8(7)\\  		
 E &2015-11-24& 57351.4182  &  57351.4232    &   --        &  0.419 &1.39 ($V$~) & $>$35 &$>$17.4(9)\\
 E &          & 57351.4176  &  57351.4238    &   --        &  0.420 &0.19 ($I_C$)& $>$35 &$>$10.1(9)\\
 F &2016-02-05&  57424.3038 &  57424.3048    &  57424.3093 &  0.917 &0.10 ($V$~)& 7.9    &1.2(3)\\
 G &          &  57424.3270 &  57424.3283    &  57424.3336 &  0.947 &0.44 ($V$~)& 9.5    &5.3(7)\\
 G &          &  57424.3278 &  57424.3289    &  57424.3327 &  0.948 &0.12 ($R_C$)& 7.1   &1.6(6)\\
      
  \tableline  
   \end{tabular}
\end{center}
\end{table*}

\begin{figure}[t]
\includegraphics[width=\columnwidth]{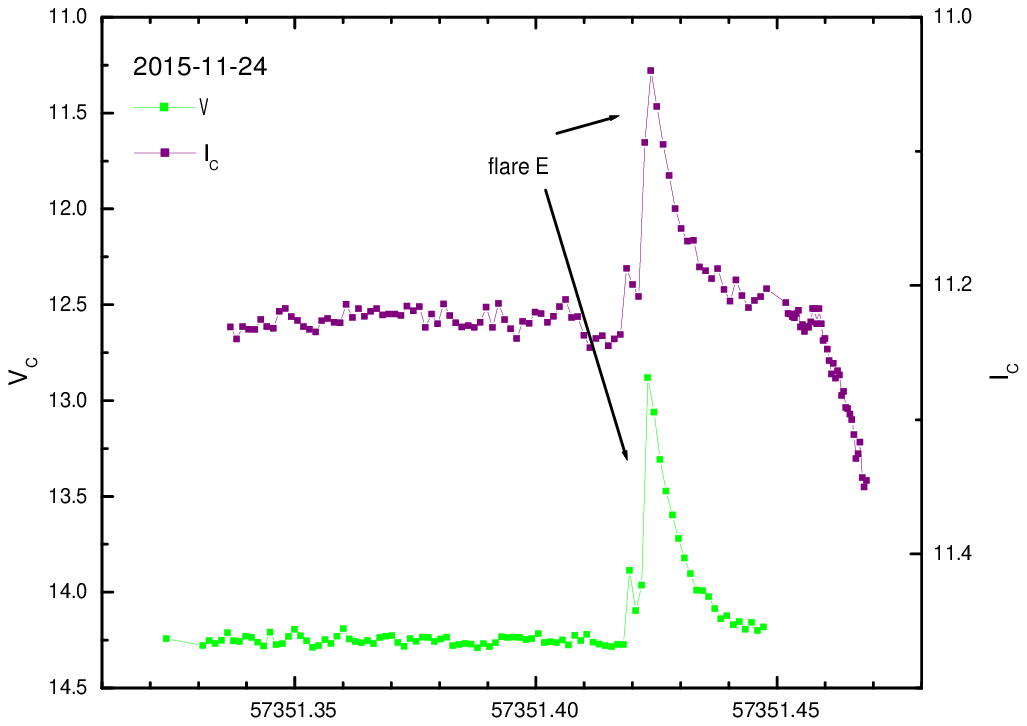}  
\includegraphics[width=0.908\columnwidth]{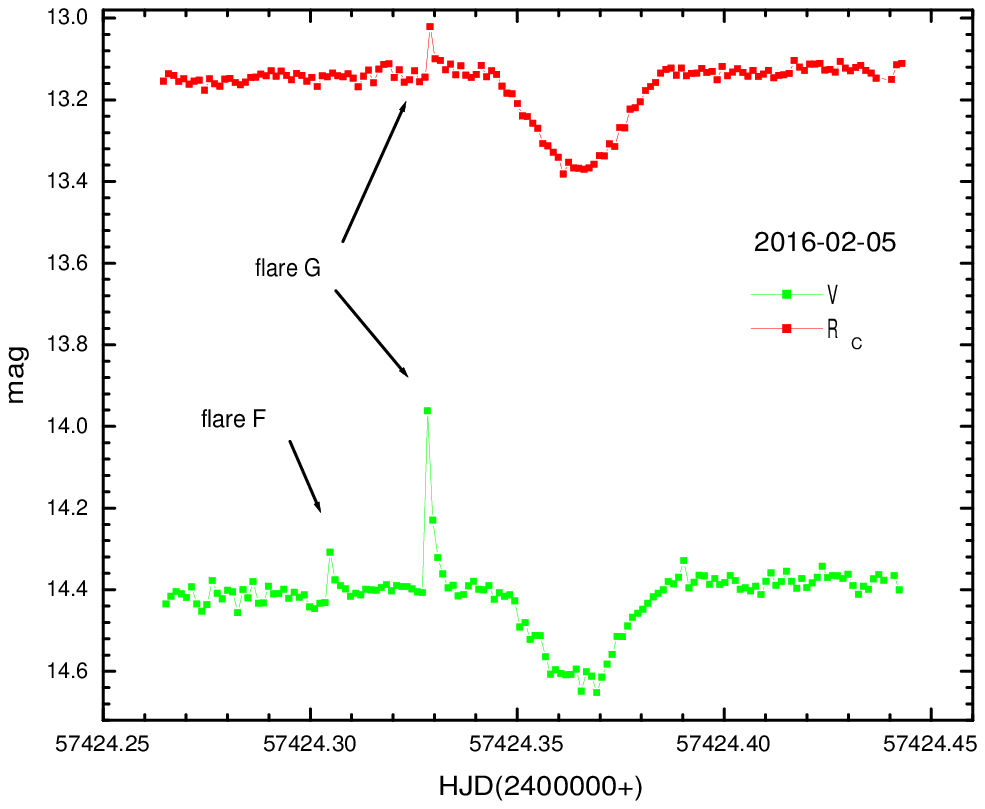}  
\caption{Flares detected in $V, I_C$ (top) and $V, R_C$ (bottom) at 2 different nights. For the parameters of the 
flares see, Table~\ref{tbl:3}. }
\label{fig:04}
\end{figure}

GJ~3236 was selected as one of the targets of The Dwarf Project \citep{pribulla2007}, which is 
focused at detection of sub-stellar companions to low-mass (composed of late-type, sD or/and WD components) detached 
eclipsing binaries using minima timing. Our observations were from the beginning focused mostly on determination of  
primary minima times and we did not gather regularly all light curves. This considerably restricted our chance 
to detect any flares. So we do not list these observations in Table~\ref{tbl:2}. 

We used two criteria for flare detection: (1) a flare should last for several minutes - at least 3 times longer than 
the used exposure time for a CCD image, (2) at least two successive data points on the light curve are brighter (within 
their errors) than the quiescent surrounding of the fitted light curve plus 3$\sigma$ level. The light curve around 
potential flare was fitted by the 3$^{rd}$-order weighted polynomial function. Weights were selected as 1/$e_{i}^2$, 
where $e_i$ are magnitude errors of individual points determined from photometry. The start and the end of the flare was 
determined visually according to the fitted light curve, brightnennig $\Delta$m was calculated as the difference 
between the maximum brightness and the fitted light curve.

The first flare event (labeled A) was observed on October 31, 2015, with the maximum at the phase of $\sim$0.38 with 
the amplitude of about 0.04 mag in the $R_C$-filter and the duration of about 6 minutes. It is a disputable flare, 
because it was on the edge of our detection criteria, which is caused by quite poor night quality, but brightening is 
clearly visible on the light curve (see Fig~\ref{fig:03}, top) The next night no flares were detected within the 
observational errors and detection criteria.

The evident flare (B) was detected just after the primary minimum on November 3, 2015, (see Fig~\ref{fig:03}, middle) 
at the phase of 0.023. We cannot exactly determine when this flare started, although we can see no distortion of 
the minimum light curve.  It means that the flare started probably shortly at the end of the minimum or just 
after it. The amplitude reached 0.24 mag in the $R_C$-filter and the flare lasted more than 10 minutes.

Another two flares (C and D) occurred on November 5, 2015, (see Fig~\ref{fig:03}, bottom) at the phase of 0.734 resp. 
0.814. Duration of the first one was more than 46 minutes with the amplitude of 0.10 mag in the $R_C$-filter. The 
second one was a short-term (6.7 minutes) and low-amplitude (0.03 mag in the $R_C$-filter) flare.

On November 24, 2015, a strong flare (E) was detected simultaneously in $V$ and $I_C$ passbands (see Fig~\ref{fig:04}, 
top). The flare lasted more than 35 minutes at the phase of 0.421 and its amplitude eas 1.39 mag, resp. 0.19 mag in 
$V$, resp. $I_C$-filters.

Another two flares (F and G) were observed on February 5, 2016. The first one was detected only in the $V$-filter at the
phase of 0.917, lasted $\sim$8 minutes with the amplitude of 0.1 mag. The second flare was detected  in the $V$ and 
$R_C$-filters at the phase of 0.947. The flare in $V$ lasted more than 9 minutes and the amplitude was 0.44 mag., while 
the $R_C$ flare lasted 7 minutes with the amplitude of 0.12 mag. 

No flares were detected during our last observation on March 19, 2016. 

We tried to roughly determine the energy emitted by the flares. We assumed bolometric quiescent luminosities of the 
components determined by \cite{irwin2009} ($L_1^{bol}$=0.016$L_\odot$, $L_2^{bol}$=0.009$L_\odot$) and the Planck law 
of the energy distribution of the flare emission. Quiescent luminosity of a star with the radius $R$ in the specific 
filter $\mathfrak{F}$ can be estimated as: 
\begin{equation}
	L(\mathfrak{F}) = \int_{0}^{\infty} 4\pi R^{2} S_\mathfrak{F}(\lambda) F(\lambda) d\lambda,
\end{equation}
where $S_\mathfrak{F}(\lambda)$ is the transmission curve of the filter and $F(\lambda)$ is the out-of-flare flux 
of the star at the wavelength $\lambda$. The total emitted energy by the flare in the filter is calculated by 
\begin{equation}
 E(\mathfrak{F}) = L_{1+2}(\mathfrak{F}) \int_{t_B}^{t_E} (10^{-0.4 \Delta m(t)} -1) dt,
\end{equation}
where we integrated from the beginning $t_B$ to the end $t_E$ of the flare, $L_{1+2}(\mathfrak{F})$ denotes the total 
luminosity of the binary and $\Delta m(t)$ is the magnitude evolution of the flare with time. Properties of 
all detected flares are summarized in Table~\ref{tbl:3}.

\section{Discussion and conclusion}
\label{disscussion}
Our observations of GJ~3236 showed that this low-mass system is a very active binary. The solution of our 
$R_C$-light-curve confirmed previously determined values for the components' radii and masses by \cite{irwin2009}. But 
the light curve has significantly changed since Irwin's observations. This can be explained by the evolution of the 
spot(s) on the components. We found that there probably exists one large spot or a large area of many smaller spots 
close to the pole of the primary component. The radius of this spot or area is about 1.5 times larger than in 2008 
(the year of MEarth observations). Another spot should exist on the primary, or on the secondary component.
Unfortunately, we cannot decide without additional, mainly spectroscopic observations, which spot's solution is 
correct. 

Such big spots suggest enormous magnetic activity of the system and one can expect flare events. We have 
photometrically monitored  GJ~3236 for more than 65.7 hours and detected 7 flare events, which revealed a flare rate of 
about 0.1 flares per hour assuming an even time distribution of the flares. This is similar to flare rate of DV Psc, 
0.082 flares per hour \citep{pi2014}, and about two times more than in other similar systems CU~Cnc, 0.05 flares per 
hour \citep{qian2012} and CM Dra, 0.057 flares per hour \citep{nelson2007}.

The flares are detected mainly in the short wavelength passbands ($U$ and $B$). The fact that we observe flares in 
$R_C$ and even $I_C$-filters suggests their high energy. The energy determined in the specific passbands as given in 
Table \ref{tbl:3} should be considered as the lower limit, mainly because of the uncertain time interval of the flare, 
its amplitude and out of flare level of the light curve. We also do not take into account spots for luminosity 
determination of the components. Moreover detection of a flare depends not only on its energy, but also on its position 
with respect to the observer. No information about positions of flares can be obtained from the photometry. Positions 
of all detected flares are uncertain, except for the most energetic flare E, which probably occurred on the primary 
component, given the phase when it was observed and the system geometry. We should conclude that 
more realistic values of the emitted energy are more than one order higher than the values given in Table 
\ref{tbl:3}. According to the review of \cite{pettersen1989} we can estimate bolometric flare energy to 
be $\sim10^{27}$~J.

Photometric precision and the time resolution of our observations do not allow us to observe weak flaring events. Only 
two such low energy flares in $R_C$ (flares A and D) and one in $V$ (flare F) passbands were detected using our 
detection criteria. Flare G can be also regarded as a low energy short-term flare. These types of flares last 
not more than a few, up to 10 minutes and no signs of any structure on the LCs are visible. This is only a few points 
over detection limit.

More energetic flares C and E lasted several tens of minutes and the shapes of their LCs are more complicated. On the 
rising part of the LCs we can detect small brightness decrease (flare E) or a short plateau (flare C) followed by steep 
brightening. The decreasing part of the LC could be approximated by an exponential decay. \citet{davenport2014} showed, 
from analysis of Kepler observations of GJ~1243, that such a complex behavior of flares is very common for long lasting
flares with duration more than 50 minutes. That is in agreement with our observations, although our determination of 
flares duration is uncertain. This complex behavior of flares could be connected with physical processes linked with 
active regions and magnetic energy dissipation or it can be explained by a random superposition of different flares 
from separate regions. To reveal the rate, properties and nature of the complex, as well as low-energetic flares, one 
need to observe them in at least two passbands with high time resolution. 

\acknowledgments
 This work has been supported by the project APVV-15-0458 of the Slovak Research and Development Agency. M.V. would 
like  to thank the project VEGA 2/0143/14.


\end{document}